\documentclass[aps,prb,twocolumn,superscriptaddress]{revtex4-2}

\usepackage{graphicx}
\usepackage{dcolumn}
\usepackage{bm}
\usepackage{amsmath}
\usepackage{gensymb} 
\usepackage[colorlinks,linkcolor=blue,anchorcolor=blue,citecolor=blue,urlcolor=blue,filecolor=blue,menucolor=blue,runcolor=blue]{hyperref}
\usepackage{ulem}

\begin{document}

\title{AC magnetic measurements with a self-oscillating LC circuit and its application to university education}

\author{Harshit Agarwal}
\affiliation{Department of Physics, Missouri University of Science and Technology, Rolla, Missouri 65409, USA}

\author{Oleksandra Uralska}
\affiliation{Department of Physics, Missouri University of Science and Technology, Rolla, Missouri 65409, USA}

\author{Jasmin Billingsley}
\affiliation{Department of Physics, Missouri University of Science and Technology, Rolla, Missouri 65409, USA}

\author{Maxim Yamilov}
\affiliation{Department of Physics, Missouri University of Science and Technology, Rolla, Missouri 65409, USA}

\author{Hyunsoo Kim}\email{hyunsoo.kim@mst.edu}
\affiliation{Department of Physics, Missouri University of Science and Technology, Rolla, Missouri 65409, USA}
\affiliation{Materials Research Center, Missouri University of Science and Technology, Rolla, Missouri 65409, USA}

\date{\today}

\begin{abstract}
Understanding the magnetic properties of matter plays a key role in materials physics.
However, university education on fundamental magnetism is limited to a theoretical survey because of the lack of appropriate apparatus that can be applied for laboratory courses at the undergraduate level.
In this work, we introduce an AC magnetometer based on the Colpitts self-oscillator with an inductor coil as a probe.
We show that this type of self-oscillator can be adopted in a typical university laboratory course to learn the principles of magnetic measurement and to understand the fundamental magnetism of matter.
We demonstrate the exceptional stability of the circuit with a working frequency range of 10 kHz to 10 MHz and excellent performance to detect a diamagnetic signal from a superconductor at cryogenic temperature.
\end{abstract}



\maketitle

\section{Introduction}
Magnetism is a principal property of matter and is frequently investigated for materials characterization, which has become common practice in research and academic institutions as well as in various industry sectors.
However, the magnetism of matter has not been sufficiently emphasized in university education and remains largely theoretical. 
Furthermore, experimental methods in magnetism are limited in university laboratory courses, resulting in an insufficient understanding of magnetism for undergraduate students.
Here, we offer a simple implementation of a magnetic measurement apparatus in an undergraduate laboratory, which would help students understand basic concepts of material magnetism and the principles of magnetic measurements.

The most fundamental magnetism includes paramagnetism and diamagnetism \cite{Blundell2001}, typically expected in a metal and an insulator, respectively.
While the orbital motion of core electrons contributes to diamagnetism, the paramagnetism arises mainly from electrons' intrinsic spin magnetic moments, resulting in the Pauli paramagnetism from the itinerant electron or Curie-Weiss paramagnetism from the localized magnetic moment \cite{Anderson1961, Anderson1978}.
Magnetic moments in a paramagnetic material can spontaneously order below a critical temperature, becoming a ferromagnet with finite net magnetization $M$ or an antiferromagnet where $M=0$.

The central quantity of research on materials magnetism is magnetic susceptibility $\chi=dM/dH$, typically obtained from the measured $M$ in a finite magnetic field $H$.
Its sign, magnitude, and dependence on temperature and field provide comprehensive information about the magnetic state of a solid. 
The principle of basic magnetic measurement is the linear response of the induced magnetization due to an applied magnetic field, $M=\chi H$, at a given time and position, and the observation of $M$ allows the determination of $\chi$ for a given $H$.

Magnetic measurement techniques detect magnetization in various ways \cite{Mugiraneza2022}.
The superconducting quantum interference device (SQUID) has become a standard method to measure DC magnetization \cite{Fagaly2006}.
In this technique, a magnetized sample passes through an inductor where the magnetic flux change induces a subtle electric current, which is measured by using a Josephson junction device.
Although this technique is one of the most precise methods, it is expensive for an average educational institution.
Furthermore, the dynamic nature of magnetism in the sample cannot be studied because the measurement typically takes a few seconds.
On the other hand, AC magnetic susceptibility with varying frequencies enables the investigation of the dynamic nature of magnetism and also allows studies of complex magnetic susceptibility that depends on the frequency of measurement \cite{Nikolo1995, Topping2019}.

A conventional AC magnetic susceptibility technique requires two concentric coils. The primary outer (or driving) coil is used to generate an AC magnetic field and therefore induces $M$ in the sample inside the secondary (or pick-up) coil, which detects the oscillating $M$.
The amplitude and phase of the induced current in the secondary coil are converted to AC magnetic susceptibility \cite{Gomory1997}.
In cryogenic applications, the sample and both coils are located inside a research refrigeration system, and long electrical leads such as twisted pairs and coaxial cables are introduced between the coils and a measurement device, typically a lock-in amplifier.
The type of long leads fundamentally limits the maximum frequency and precision because of a sizable background inductance \cite{Clover1970}.

When exceptional precision is required in an AC magnetic measurement, the inductor-capacitor ($LC$) self-oscillator can be employed by using an inductor coil as a magnetic probe.
The technique utilizing the $LC$ self-oscillator is a versatile experimental tool for the investigation of magnetic properties, and they have been employed for various scientific measurements. 
A cryogenic application of the $LC$ self-oscillator offers high sensitivity because it intrinsically eliminates the large background signal from long coaxial cables \cite{Clover1970}. 
Among such techniques, the tunnel diode-based $LC$ self-oscillator (TDO) has been successful in frequency ranges of $\sim 10$-$100$ MHz \cite{Clover1970, Degrift1975, Vannette2008, Kim2018rsi} for magnetism \cite{Vannette2008} and superconductivity \cite{Carrington1999, Kim2018rsi}.
Measurements with an $LC$ self-oscillator are typically performed in the frequency domain, and the accurate determination of the frequency shift is crucial, which is directly related to the magnetic susceptibility of a specimen, and 0.001 ppm precision can be readily achieved \cite{Degrift1975, Kim2018rsi}.

Although a cryogenic TDO technique has been successful, the TDO signal is generally unstable at room temperature.
It also has a narrow working voltage range of, for example, about 150 mV \cite{Kim2018rsi}.
Furthermore, the tunnel diode is a delicate component with a narrow $p$-$n$ junction width \cite{Esaki1958} and is prone to electrostatic shock that is often unpredictable.
Therefore, it is not suitable for school education in typical laboratory courses.
We found the transistor-powered Colpitts self-oscillator \cite{Colpitts, Rohde2016} can be adopted for university magnetism education by integrating it into a lab course or an independent study.
In this work, we report details of the development of the Colpitts self-oscillator as a magnetic probe and its potential application to magnetism and circuit education at the university level.

\section{Experimental methods}

For electrical circuits, we used commercial passive components and various bipolar junction transistors.
We tested the circuit with components from various sources and found little or no difference in performance.
Low-power tunnel diodes are not widely available and relatively expensive, and we tested the tunnel diode oscillator (TDO) with an inexpensive tunnel diode with the peak current greater than 1 mA, which can be easily obtained from online retail stores.
TDO was built on a prototyping, printable circuit board (PCB).
For the Colpitts oscillator, we used both a plugable circuit board (breadboard) and various PCBs.
A common benchtop power supply, oscilloscope, waveform generator, and frequency counter were used for measurement and data acquisition.

\section{Tunnel diode based $LC$ self-oscillator}

\begin{figure}
    \centering
    \includegraphics[width=1\linewidth]{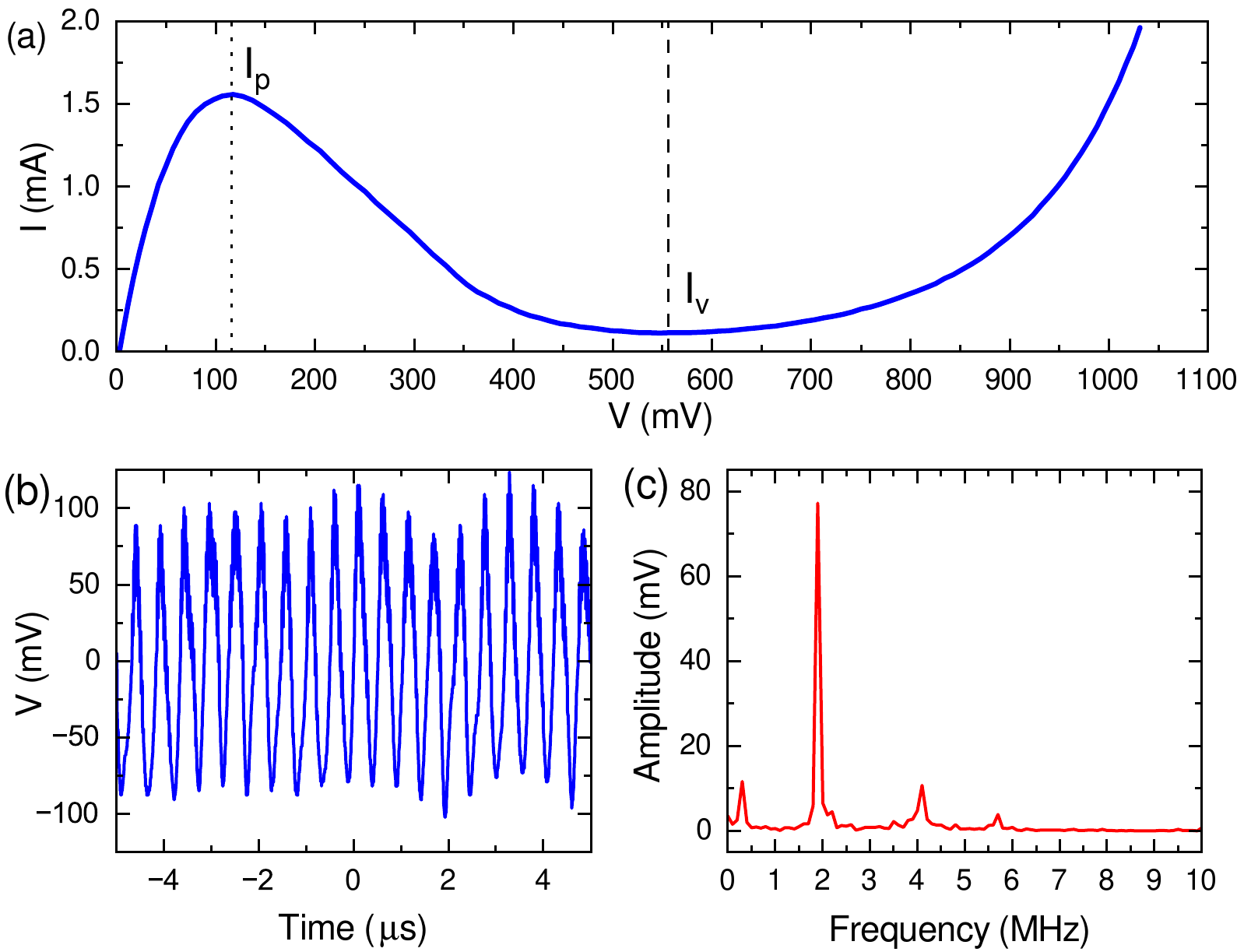}
    \caption{(a) $I$-$V$ curve of a tunnel diode. (b) Oscillatory signals from the tunnel diode circuit. (c) Fast Fourier transform of the signal in panel (b).}
    \label{fig:tdo}
\end{figure}

The tunnel diode was discovered by Leo Esaki in 1958 \cite{Esaki1958}. It is a $p$-$n$ junction diode comprised of heavily doped semiconductors, which results in substantial overlap of the valence and conduction bands of the $p$ and $n$ types, respectively \cite{Streetman1990}.
Such overlap of the bands with a narrow junction width of about 15 nm allows a quantum tunneling current even without an applied bias voltage \cite{Esaki1958}.
A weak DC voltage promotes the current up to the so-called peak current ($I_p$). However, further increase of the voltage results in the reduction of the tunneling cross-section and exhibits negative differential resistance as shown in Fig. \ref{fig:tdo}(a).
Therefore, the voltage-dependent current, or $I$-$V$ curve, can be qualitatively represented by a cubic function, $I(V)\sim V^3$. 
The cubic current function can generate a stable limit cycle that can be analytically explained with the van der Pol equation \cite{Pol1934}, a special case of the Lienard nonlinear second-order differential equation \cite{Jenkins2013}. 
Scientific applications of the tunnel diode oscillator date back to 1970, and R. B. Clover and W. P. Wolf successfully employed a TDO technique for the measurement of magnetic susceptibility \cite{Clover1970}. 
In later years, Craig T. Van Degrift made a significant improvement and successfully achieved 0.001 ppm precision measurements \cite{Degrift1975}.

Self-oscillation in a TDO circuit can be easily demonstrated even without the $LC$ tank circuit.
Due to the presence of undesired reactive components, including the intrinsic capacitance of the tunnel diode and parasitic inductance in the circuit, an oscillation easily occurs in a simple series circuit of a resistor and a tunnel diode.
Figure \ref{fig:tdo}(b) displays an example of an oscillation that occurs when the voltage, $V = 170$ mV, is applied to a tunnel diode with an $I$-$V$ curve shown in Fig. \ref{fig:tdo}(a). In Fig. \ref{fig:tdo}(c), A fast Fourier transform of the oscillation signal reveals a main frequency of 2 MHz with a few minor peaks, which are likely the ambient noise.

%
\section{Colpitts $LC$ Self-oscillator}

\subsection{Circuit Design}

\begin{figure}
    \centering
    \includegraphics[width=0.8\linewidth]{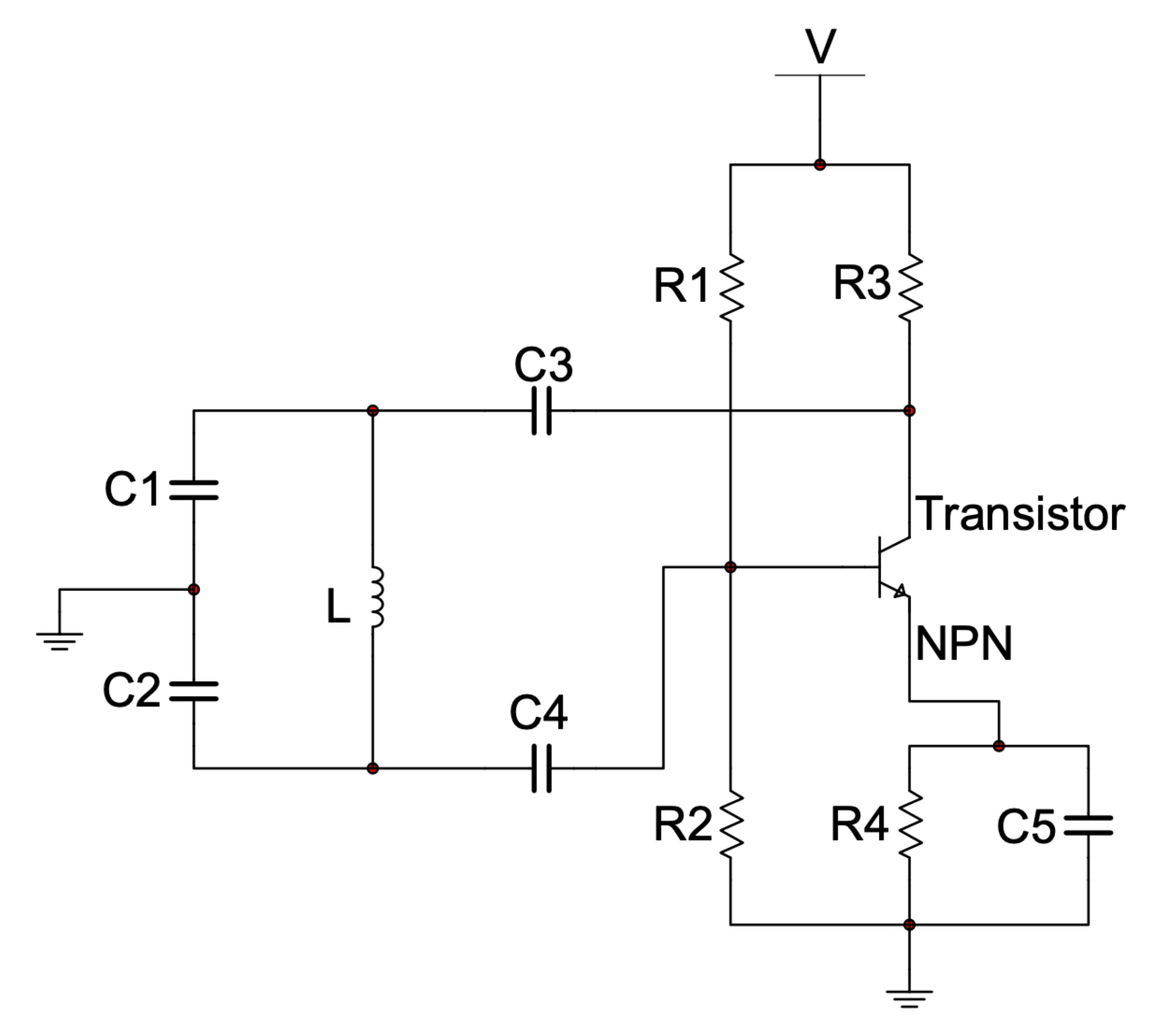}
    \caption{A circuit diagram of the Colpitts oscillator powered by a NPN bipolar junction transistor. $R_1=33$ k$\Omega$, $R_2=10$ k$\Omega$, $R_3=R_4=1$ k$\Omega$, $C_1=4.7$ nF, $C_2=47$ nF, $C_3=C_4=470$ nF, $C_5=220$ nF. The oscillation frequency is determined by the tank circuit, composed of $C_1$, $C_2$, and $L$. The minimum required bias voltage $V$ is typically $\sim 3$ V for given circuit components.}
    \label{fig:colpitts}
\end{figure}

The original Colpitts self-oscillator was powered by a vacuum tube \cite{Colpitts}. Since the introduction of the transistor, a number of variations have been developed \cite{Rohde2016}. 
In our work, we used the simplest form with the minimum number of necessary elements, as shown in Fig. \ref{fig:colpitts}. 
The electrical circuit of the Colpitts oscillator (CO) consists of two main parts, separated by two coupling capacitors, $C_3$ and $C_4$.
The left and right sides are respectively the $LC$ tank and the amplification circuits.
In the presence of reactance components, both intrinsic and parasitic, special precautions need to be taken to achieve the coherent feedback mechanism.
The critical factors for a self-oscillating circuit are (i) sufficient amplification and (ii) phase matching between the original and amplified signal.

The key component on the right-hand side is a bipolar junction transistor (BJT). 
The multiresistor circuit is constructed to properly power the transistor. 
$R_1$ and $R_2$ comprise a voltage divider and set the bias voltage to the base of the transistor. 
$R_3$ and $R_4$ determine the gain of the DC voltage.
The $C_5$, which is introduced in parallel to $R_4$, is necessary to maintain the oscillation generated by the $LC$ tank circuit (left side).
Therefore, the gain of the AC signal is determined by the ratio of $R_3$ and the internal resistance of the transistor (emitter leg). 
Keeping the voltage-divider bias ratio and tank capacitors ratio constant, the amplification depends on the ratio of the resistor connected to the collector and the internal resistance of the transistor. 

The main characteristics of CO include the use of two capacitors ($C_1$ and $C_2$), separated by a grounding, in the $LC$ tank circuit.
When the oscillator circuit is powered by a DC voltage, self-oscillation occurs with a resonance frequency $f_0$;
\begin{equation}\label{eq:fcolpitts}
2 \pi f_0 = \frac{1}{\sqrt{LC}},~C=\frac{C_1 C_2}{C_1 + C_2}.
\end{equation}
A fraction of the oscillatory signal in the tank circuit determined by $C_2/C_1$ (typically $\sim 0.1$) is superimposed on the dc current at the base of the BJT through $C_4$, and the signal is amplified in the emitter before coherently returning to the $LC$ tank circuit through $C_3$.

\subsection{Performance: stability and noise level}

\begin{figure}
    \centering
    \includegraphics[width=1\linewidth]{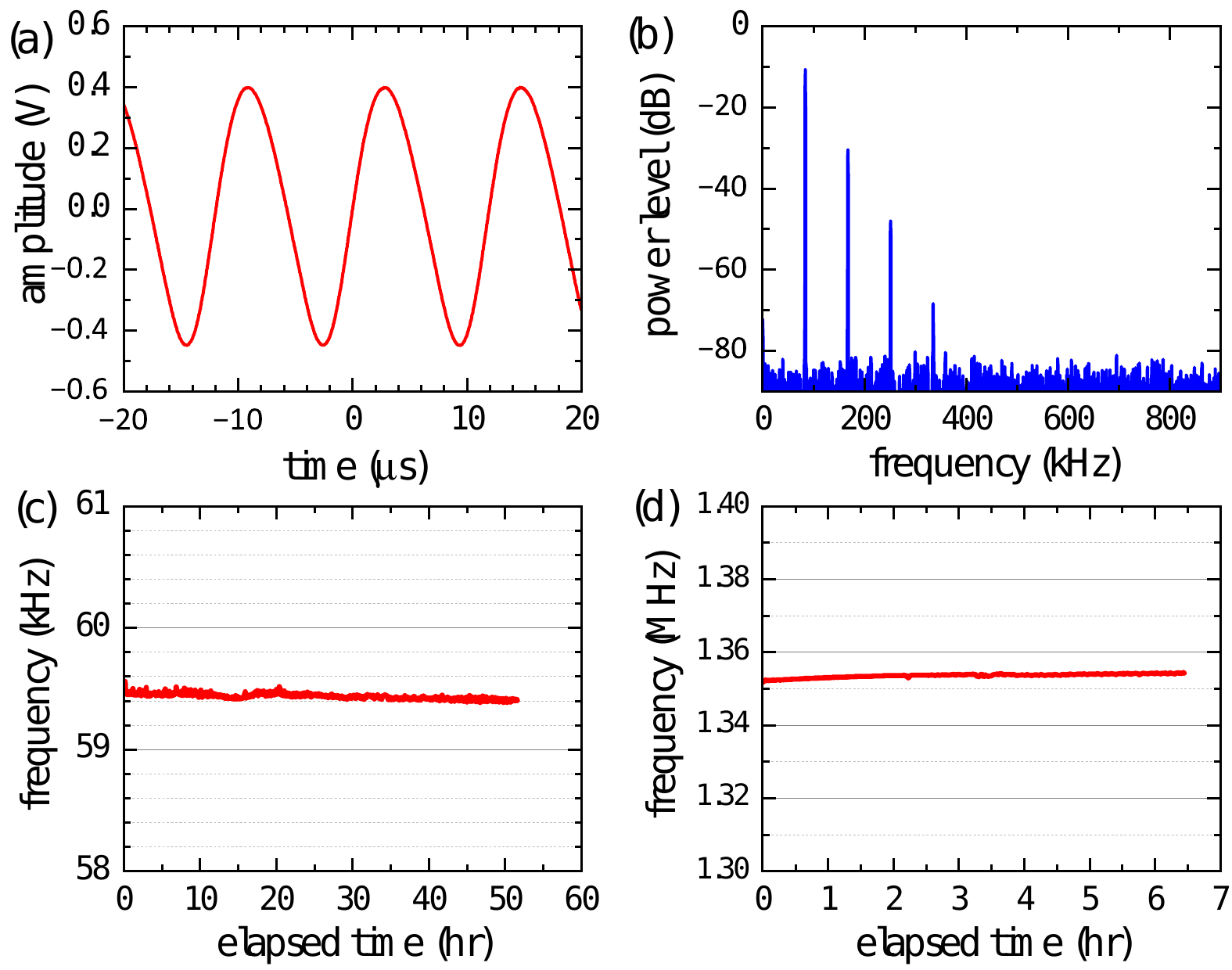}
    \caption{Performance of a Colpitts oscillator, powered by a bipolar junction transistor. (a) Typical waveform and (b) power spectrum of $f_0\approx 83$ kHz. Stability, drifting over time, and noise level for (c) 59.5 kHz oscillator and (d) 1.346 MHz oscillator.}
    \label{fig:oscsignal}
\end{figure}

We designed various COs and tested their performance using commercial resistors, capacitors, inductors, and a transistor. 
Every component was plugged into a breadboard.
The CO under test was powered at a DC voltage between 3 V and 6 V.
The oscillation frequency slightly depends on the voltage value within 1\%.

Figure \ref{fig:oscsignal} shows the output signal from various COs. 
In panel (a), the waveform appears in a nearly sinusoidal shape with a period of about 12 $\mu$s that corresponds to about 83.3 kHz.
Panel (b) shows the power spectrum of the oscillation signal, exhibiting a strong peak of fundamental frequency of 83 kHz while harmonics up to the 4th are present. 
The strength of the second peak with respect to the primary one is substantially attenuated.
Although the strength of the harmonics can be suppressed by optimally tuning the circuit components, it is acceptable for most of the applications, in particular for measurements in the frequency domain. The Colpitts self-oscillator with frequencies ranging from 10 kHz to a few MHz can readily be achieved on a breadboard, and the oscillators generally perform remarkably well.

\begin{table*}
    \centering
    \begin{ruledtabular}
    \begin{tabular}{c|ccccc}
        Oscillator & Mean (Hz) & Std. Dev. (Hz) & Min (Hz) & Max (Hz) &  Max $-$ Min (Hz) \\
        \hline \hline
        No. 1 & $5.944\times 10^4$ & $22 $ & $5.939\times 10^4$ & $5.956\times 10^4$ & $170$ \\
        No. 2 & $1.352\times 10^6$ & $5.100 \times 10^{2}$ & $1.352\times 10^6$ & $1.354\times 10^6$ & $2.750\times 10^3$ \\
    \end{tabular}
    \end{ruledtabular}
    \caption{Performance of two Colpitts oscillators: stability and noise level at room temperature}
    \label{tab:perf}
\end{table*}
To quantify circuit performance and signal quality, we tested two Colpitts oscillators with the frequencies of 50 kHz and 1.35 MHz.
The frequency of each oscillator was recorded for an extended period of time, for over 50 hours and 6 hours for 50 kHz and 1.35 MHz oscillators, respectively.
The recorded frequencies are shown in Fig. \ref{fig:oscsignal}(c, d) as a function of time. 
The noise levels for the 50 kHz and 1.35 MHz oscillators are respectively 0.003$f_0$ and 0.002$f_0$.
The detailed information of each circuit is summarized in Table \ref{tab:perf}.

\subsection{Higher frequency Colpitts oscillators: plugged (breadboard) vs. soldered (PCB) }

\begin{figure}
    \centering
    \includegraphics[width=1\linewidth]{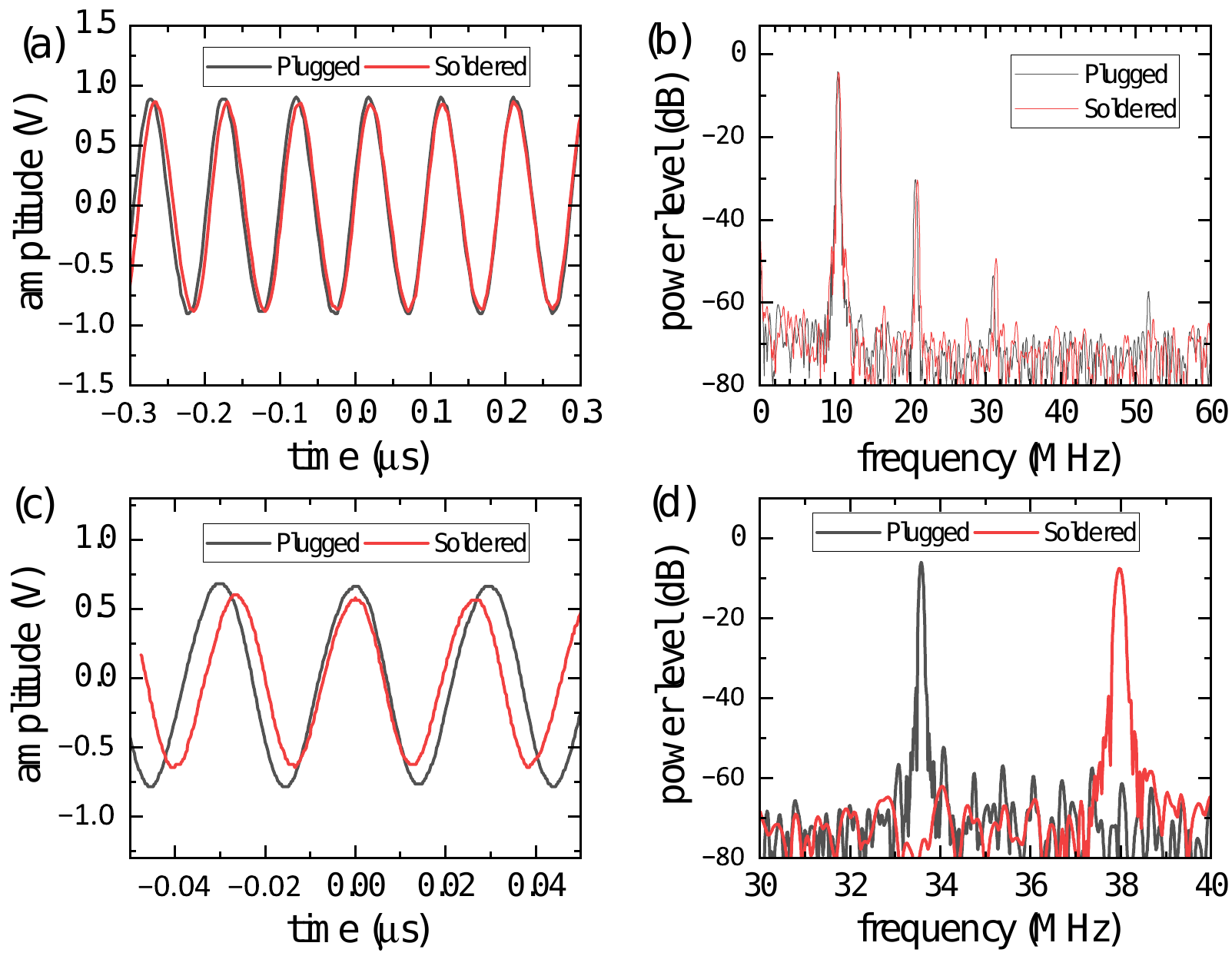}
    \caption{Performance of the Colpitts oscillators on bread board (black) and PCB (red). (a) Waveform and (b) power spectrum from a 10 MHz oscillator are shown, where similar signals are apparent. However, for a 35 MHz oscillator (a) waveform and (b) power spectrum show substantially different results between plugged and soldered circuits.}
    \label{fig:MHz}
\end{figure}

Whereas the Colpitts oscillator on the breadboard works well up to a few MHz, we noticed a significant degradation of the signal quality as well as an inconsistent frequency yield from the intended $f_0$ with $L$ and $C$ values for the frequency range above $\ sim$10 MHz.
We designed 10 MHz and 35 MHz Colpitts oscillators and, first, tested them on the breadboard by recording the waveform and power spectrum.
We subsequently built the same oscillators on prototype PCBs using the same components used for the breadboard counterpart. 
Figure \ref{fig:MHz} shows the waveforms and power spectra in the 10 MHz and 35 MHz Colpitts oscillators in the upper (a, b) and lower (c, d) panels, respectively. 
The waveforms of the 10 MHz oscillator shown in panel (a) show minor differences, and the soldered oscillator on a PCB shows a slightly higher frequency.
We attribute the noticeable difference to the possible presence of parasitic reactances of the breadboard and inconsistent contact quality.
The frequency shift in the 35 MHz PCB oscillators is evident in the power spectrum, as shown in panel (d).
Therefore, the upper frequency limit for the plug-in CO is about 10 MHz.

\subsection{Application of Colpitts oscillator to magnetic measurements}\label{sec:appmag}

The Colpitts self-oscillator with a solenoid-type inductor coil can be used as a magnetic probe to understand the magnetism of matter. 
For the simplest demonstration of magnetic sensitivity, a permanent magnet can be brought into the vicinity of the inductor coil, which causes a change in the oscillation waveform where the amplitude and frequency are affected.
When a sample is inserted inside the coil, the frequency shift is directly related to its magnetism (see the Appendix).
The TDO self-oscillator has been utilized for magnetic materials \cite{Vannette2008} and superconductors \cite{Kim2018rsi}.
In principle, the Colpitts oscillator can be applied to similar research activities, and we demonstrate its potential for superconductivity research as well as university education by using a dipper-type and a polycrystalline YBa$_2$Cu$_3$O$_{7-\delta}$ (YBCO) sample \cite{Wu1987}.

\begin{figure*}
    \centering
    \includegraphics[width=0.8\linewidth]{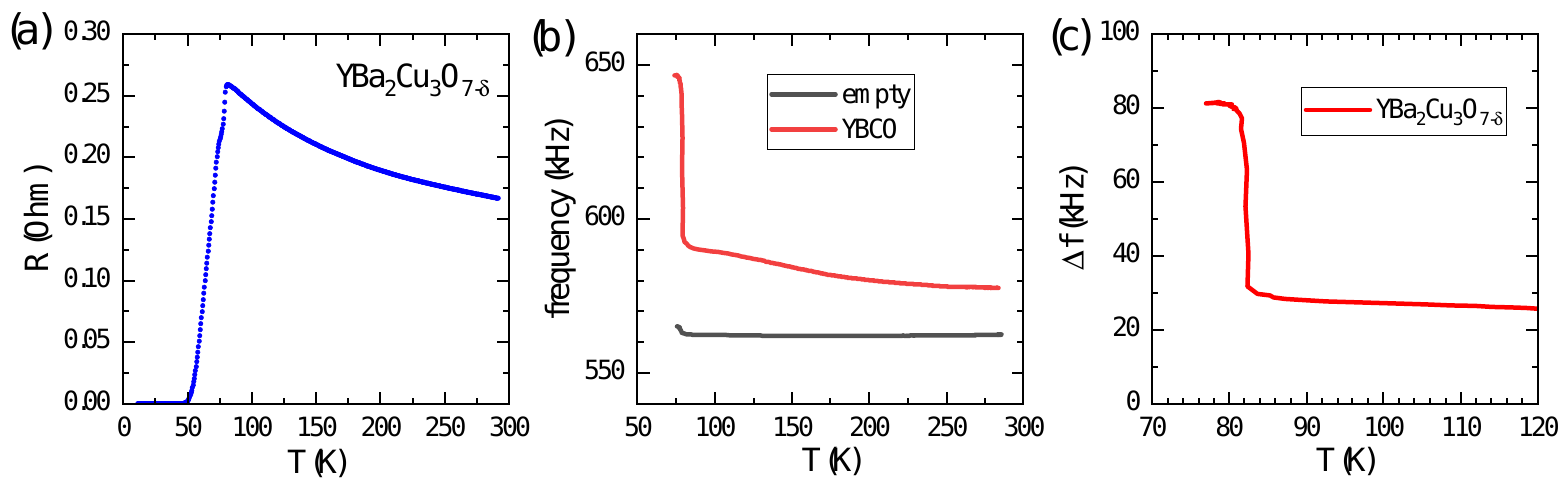}
    \caption{(a) Electrical resistance vs. temperature in a polycrystalline YBa$_2$Cu$_3$O$_{7-\delta}$ (YBCO) sample. (b) temperature dependent frequency of an empty coil (black) and with a YBCO sample (red) (c) $\Delta f(T)$ due to the magnetic susceptibility of the YBCO sample.}
    \label{fig:ybco}
\end{figure*}

Figure \ref{fig:ybco}(a) shows the temperature dependence of the electrical resistance $R(T)$ in a YBCO sample.
We used a standard four-probe method with a lock-in amplifier.
The sample was cooled in a commercial closed-cycle refrigerator with a base temperature of 10 K.
$R(T)$ is non-metallic between about 80 K and room temperature. 

$R(T)$ exhibits a sudden drop below 80 K, signaling a superconducting phase transition. 
However, $R(T)$ remains finite between 50 K and 80 K.
The broad superconducting transition of about 30 K is a reminiscence of the original discovery of the superconducting state in the high-$T_c$ cuprate superconductor \cite{Bednorz1986}.
We note that the hallmark of superconductivity, zero electrical resistance, cannot be observed at liquid nitrogen temperature with the particular YBCO we used.

We measured $\Delta f(T)$ in a YBCO sample from the same batch using a Colpitts self-oscillator with an inductor coil that is separated from the main circuit board by a 30 cm long coaxial cable along a stainless steel tube.
The resonance frequency of this oscillator is $f_0 \approx 562$ kHz.
For this activity, a long stainless steel tube that separates the $LC$ circuit from the main part of the Colpitts oscillator is slowly lowered into liquid nitrogen in a thermally insulated container such as a double-walled flask, commonly known as a dewar. 
The temperature of the coil and sample was adjusted by the distance control relative to liquid nitrogen. 
The sample temperature was determined with a solid-state cryogenic thermometer, for instance, a thermocouple or a small spool of fine copper wire. The former requires a precision voltmeter with a resolution better than 1 mV. The latter method utilizes the linear temperature relationship of its electrical resistance between room temperature and liquid nitrogen. 

The temperature dependence of the background signal with an empty coil, $f_0(T)$, was determined, which is virtually temperature independent except for an uptick near 77 K, as shown in Fig. \ref{fig:ybco}(b) (black curve). 
A YBCO sample is placed in the same inductor coil after the background signal is determined, and a measurement of $f(T)$ is performed between room temperature and 77 K in liquid nitrogen.
Figure \ref{fig:ybco}(b) shows $f(T)$ from the YBCO sample (red curve).
The overall frequency signal increases with the presence of the YBCO sample by about 15 kHz at room temperature.
With decreasing temperature, $f(T)$ gradually increases before it sharply jumps around 82 K by more than 50 kHz, which is almost 10\% of $f_0$.
The background subtracted curve, $\Delta f = f_\textmd{\tiny sample}(T)-f_0(T)$, is shown in panel (c) with temperatures below 120 K.
The sudden increase in $\Delta f$ is consistent with the onset of diamagnetism in the superconducting state of the YBCO.

\subsection{Cryogenic application}

One of the most important scientific applications of the $LC$ self-oscillator technique is the superfluid investigation of superconductors at low temperatures near absolute zero \cite{Kim2018rsi,Kim2018}.
Such an application requires exceptional precision, and cryogenic TDO technique has been successfully employed \cite{Degrift1975,Kim2018rsi}.
To maximize sensitivity or signal-to-noise ratio, it is desired to implement the circuit at a cryogenic temperature to reduce Johnson–Nyquist noise \cite{Johnson1928,Nyquist1928} and physically close to the sample to minimize the reactive contribution of the coaxial cable \cite{Clover1970}.
Although the Colpitts self-oscillator can be used at cryogenic temperatures, it has not been realized to the best of our knowledge.

Understanding the temperature dependence of electrical components is a prerequisite for the implementation of a circuit at cryogenic temperatures.
In principle, passive components can be chosen from the wide variety to minimize the effect of their temperature dependence.
However, the options for active electrical components are generally limited.
Interest in the low-temperature characteristics of active components \cite{Gerritsen1951} and small-scale devices has been consistently increasing, and this knowledge has become crucial to improve their performance. The most prominent examples include a liquid nitrogen-cooled supercomputer \cite{Carlson1989}.
For cryogenic applications of TDO, the characteristic of the tunnel diode has been well understood \cite{Esaki1958, Degrift1975, Kim2018rsi}, and the tunnel diode performs well even at 22 mK \cite{Grytsenko2025}.
However, the performance of BJT at cryogenic temperatures depends on multiple factors, including construction materials, and it is generally perceived that the germanium-based BJT performs at low temperatures better than the silicon counterpart \cite{McAlister2000}. III-V semiconductors with high electron mobility are also good candidates \cite{Grahn2020}. However, most of the commercially available BJTs are made of silicon, and, therefore, the performance of the Colpitts oscillator with a common BJT at low temperatures is questionable.

\begin{figure}[b]
    \centering
    \includegraphics[width=1\linewidth]{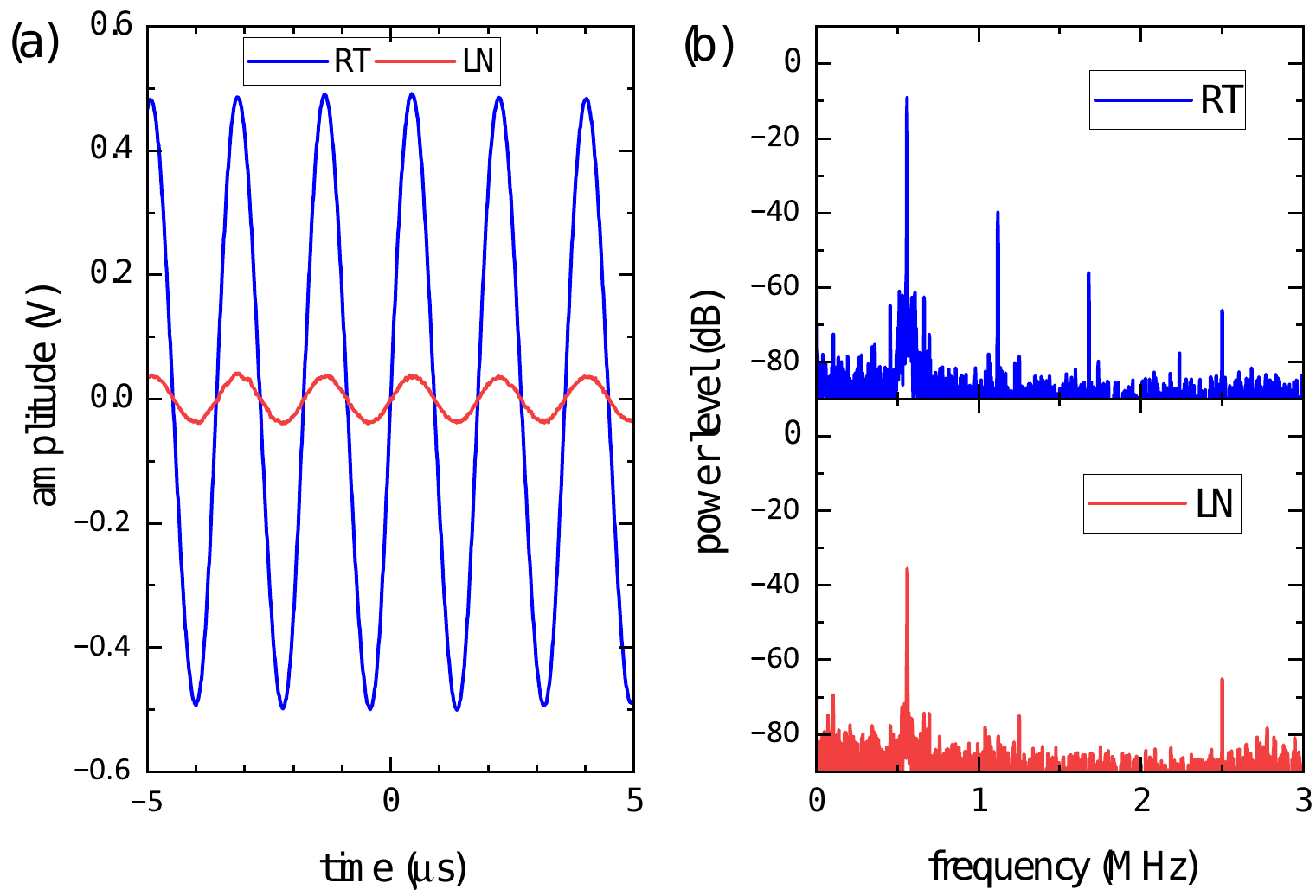}
    \caption{Cryogenic Colpitts oscillator. (a) Waveform of the oscillation signals at room temperature (blue) and liquid nitrogen (red), (b) power spectrum at room temperature (upper part in blue) and liquid nitrogen (lower part in red).}
    \label{fig:tdependent}
\end{figure}

We tested the performance of a Colpitts oscillator with a commercial silicon BJT submerged under liquid nitrogen, while the rest of the circuit is at room temperature.
We used thin, long copper wires between the transistor and the circuit board.
In Figure \ref{fig:tdependent}, we compare characteristics of the circuit with a BJT at 77 K to those of the room temperature circuit.
Panel (a) shows waveforms for both circuits, and the amplitudes are 0.5 V and 0.05 V for the room-temperature and liquid nitrogen circuits, respectively, although the oscillation signal maintains a sinusoidal waveform without apparent distortion and change in frequency at liquid nitrogen.
Their power levels are shown in panel (b).
In addition to a reduction in the amplitude of the primary frequency in liquid nitrogen, higher-order harmonics are invisible, which makes the low-temperature Colpitts oscillator potentially better in the frequency-domain measurement.
We note that a peak near 2.6 MHz appears in both tests with nearly the same power, indicating a background noise signal.

\section{Application of Colpitts oscillator for University education}
Electromagnetism is one of the most important topics in physics and serves as the foundation in various STEM fields.
While electricity and magnetism equally consist of a university course on the topic, the laboratory exercises in a classroom are mostly comprised of electricity-related topics, and most institutions cannot sufficiently provide the proper magnetism education in the laboratory. 
Needlessly to say, prompt implementation of tailored topics for the magnetism experiment is desired.

The Missouri S\&T Physics program offers independent study opportunities on topics covering magnetism basics and application to quantum materials research. and these opportunities serve as key high-impact educational practices \cite{Fischer2021}. 
The topics include fundamental concepts in magnetism and measurement techniques involving magnetic susceptibility and an $LC$-circuit, which are key to the understanding of various magnetic materials and the working principles of magnetic measurement instruments for solid-state. 
Such topics are sometimes covered in upper-level undergraduate courses, which are often too advanced for a broad spectrum of student levels, posing educational challenges. 
We tackle the issue by designing the modular research program as an exercise-based hands-on experience, which can be implemented into the existing lab courses.

Modular research programs can be created based on the Colpitts self-oscillator as a main element. 
The programs are designed to instruct the necessary knowledge for understanding electromagnetism/superconductivity and the self-oscillator technique operating in cryogenic environments. The possible modular lab activities include (i) characterization of the relevant electrical components including transistors and tunnel diode at room temperature and liquid nitrogen temperature by measuring $I$-$V$ curves, (ii) designing and building the Colpitts self-oscillator, powered by a transistor, on a plug-in breadboard and subsequently on a solderable circuit board including a printable circuit board, (iii) quality testing and fine-tuning of the oscillator with necessary theories and techniques including the spectrum analysis and fast Fourier transform, (iv) frequency dependence measurements to understand superconducting penetration depth and normal skin effect, and (v) temperature dependence measurement by dipper-type experiments for the understanding of cryogenic thermometry and various solid-state thermometers.

The culminating research experience in these independent study programs can be achieved by utilizing an inductor probe for the measurement of the relative frequency changes due to the temperature-dependent ac magnetic susceptibility in various materials. 
A dipper-type experimental probe can be utilized to measure the temperature-dependent magnetic susceptibility in a high-$T_c$ superconductor such as YBa$_2$Cu$_3$O$_{7-\delta}$ (YBCO) with $T_c \approx 93$ K \cite{Wu1987}, which can be detected with the aid of liquid nitrogen, as discussed in Section \ref{sec:appmag}. 

The modular research program for undergraduate researchers offers opportunities to be trained with practical laboratory skills and various software programs required for the lab automation of data acquisition and analysis.
Successful completion of these activities will establish (1) unique independent study programs on magnetism in quantum materials, (2) summer/winter extracurricular activities, and (3) participation opportunities for research at the forefront of quantum materials physics. 
In addition, the participating students and research advisors will establish a mentorship that will primarily offer guidance on advancing in the profession. 
Working among peers will also provide leadership, mentorship, and a greater sense of belonging to the discipline. Such outcomes will not only increase awareness of their identity but also lead to a better retention rate in the quantum STEM field.

\section{Summary}
We demonstrated that the Colpitts self-oscillator can be employed in university magnetism education because of its stability, which can be relatively easily achieved. With common commercial components, the oscillator circuit can be conveniently built on a plug-in breadboard, which exhibits excellent stability for days. When a solenoid coil is used as an inductor of the tank circuit, it shows good sensitivity as a magnetic probe, and the Colpitts oscillator can be used to study magnetism and superconductivity, making it a potential education tool for quantum materials at the undergraduate level.

\begin{acknowledgments}
H.K. is grateful to Matthew Vannette and Robert Duncan for inspiring discussion and to Sandeep Puri for the preliminary work.
H.A., J.B., O.U. are grateful for support by First Year Research Experience (FYRE) and Opportunities for Undergraduate Research Experience (OURE) at Missouri S\&T. H.K. is grateful for support from Materials Research Center (MRC) Seed Funding, M.Y. and H.K. are grateful for support from Kummer Center for STEM Education via SPARC program. H.K. thanks Missouri S\&T Seed Funding. The research was supported by NSF Award No. DMR-2418630
\end{acknowledgments}

\appendix*

\section{frequency shift and magnetic susceptibility}
The resonant frequency of an empty $LC$ tank circuit is given by
\begin{equation}
f_0=\frac{1}{2\pi\sqrt{LC}}.
\end{equation}
When a sample is placed inside the inductor $L$, the resonant frequency changes by $\Delta f$ due to the extra inductance $\Delta L$ due to the presence of the sample.
The new resonance frequency can be written as follows.
\begin{equation}
f_0+\Delta f = \frac{1}{2\pi\sqrt{(L+\Delta L)C}}=\frac{1}{2\pi\sqrt{LC}}\left(1+\frac{\Delta L}{L}\right)^{-1/2}
\end{equation}
When the sample volume $V_s$ is significantly smaller than the coil volume $V_c$, $\Delta L$ is small.
Provided $\Delta L \ll L$, we can make the following approximation.
\begin{equation}
f_0+\Delta f \approx f_0 \left(1-\frac{\Delta L}{2 L}\right)
\end{equation}
Therefore, the expected $\Delta f$ due to $\Delta L$ is given by
\begin{equation}
\Delta f \approx -\frac{1}{2} \frac{\Delta L}{L}f_0~~\textmd{or}~~\frac{\Delta f}{f_0}\approx -\frac{1}{2} \frac{\Delta L}{L}.
\end{equation}

The inductance $L$ can be expressed in terms of magnetic flux $\Phi$ and electrical current $I$, i.e., $L=\Phi/I$, 
$\Delta L$ arises due to the change in magnetic flux $\Delta \Phi$. Therefore,
\begin{equation}
\Delta L = \frac{\Delta \Phi}{I}~~\textmd{or}~~\frac{\Delta L}{ L} = \frac{\Delta \Phi}{ \Phi}.
\end{equation}

The total magnetic flux in the absence of a coil is given by
\begin{equation}
\Phi \approx H_{ac} A l n = H_{ac} V_c n.
\end{equation}
Here, $H_{ac}$ is the ac magnetic field inside the coil, $A$ is the cross-sectional area of the inductor coil, $l$ is the length of the coil, and $n$ is the number of turns per unit length.

When a sample with magnetic susceptibility $\chi$ is placed inside the coil, the sample will displace a volume with the sample volume $V_s$, resulting in $\Delta \Phi \approx M V_s n$. Here, $M$ is the magnetization of the sample, and $M=\chi H_{ac}$. Therefore,
\begin{equation}
\frac{\Delta \Phi}{ \Phi} \approx \frac{V_s}{V_c} \chi.
\end{equation}
Finally, the frequency shift due to the magnetic sample can be written as follows. 
\begin{equation}
\frac{\Delta f}{f_0}\approx-\frac{1}{2} \frac{V_s}{V_c}4\pi\chi.
\end{equation}

\bibliographystyle{apsrev4-2}

\end{document}